\documentclass[a4paper]{JHEP3}

\usepackage{amsmath}

\DeclareMathOperator{\Arg}{Arg}

\def\clap#1{\hbox to 0pt{\hss#1\hss}}
\def\mathclap{\mathpalette\mathclapinternal}
\def\mathclapinternal#1#2{\clap{$\mathsurround=0pt#1{#2}$}}

\preprint{DESY 05-145}

\title{The General Warped Solution 
with Conical Branes in Six-dimensional Supergravity}

\author{Hyun Min Lee, Christoph L\"udeling\\
	\upshape Deutsches Elektronen-Synchrotron DESY, Notkestra\ss e
	85, 22607 Hamburg, Germany\\ 
	\upshape e-mail: {\ttfamily hyun.min.lee@desy.de, christoph.luedeling@desy.de}} 

\abstract{
We present the general regular
warped solution with 4D Minkowski spacetime
in six-dimensional gauged supergravity.
In this framework, we can easily embed multiple conical branes
into the warped geometry  by choosing an undetermined holomorphic function.
As an example, for the holomorphic function with many zeroes, 
we find warped solutions with multi-branes and discuss
the generalized flux quantization in this case.}


\begin{document}
%\maketitle

%%%%%%%%%%%%%%%%%%%%%%%%%%%%%%%%%%%%%%%%%%%%%%%%%%%%%%%%%%%%%%%%

\section{Introduction}

%%%%%%%%%%%%%%%%%%%%%%%%%%%%%%%%%%%%%%%%%%%%%%%%%%%%%%%%%%%%%%%%
\noindent
%%%% Codimension-two branes and self-tuning %%%%
In recent years there has been a lot of interest in models with branes embedded in higher dimensions. 
One particular motivation is the hope of finding a solution to the notorious cosmological constant problem \cite{weinberg}. 
From this point of view, 
six-dimensional (6D) models with codimension-two branes 
are especially interesting. 
It was noticed in ref. \cite{conical} that the vacuum energy localized 
on a codimension-two brane results only in a deficit angle of a conical singularity without generating an effective cosmological constant.
This observation prompted attempts of realizing the self-tuning idea \cite{self-tuning} in the 6D framework.  

%%%% Brane solutions in 6D supergravity and other models %%%%
Specific realizations were first studied in the context of a non-supersymmetric 6D Einstein-Maxwell theory \cite{flux,navarro}. 
Background solutions with 4D Minkowski symmetry exist for any values of the brane tensions, once tuning between  the gauge flux and a positive bulk cosmological constant is imposed.
Embedding such setups in 6D gauged supergravity \cite{6dsugra1,6dsugra2} 
turns out to be advantageous for several reasons.
The so-called Salam-Sezgin supergravity \cite{salam-sezgin}  
automatically provides the necessary abelian gauge field 
and positive definite potential.
Moreover,  the tuning of the gauge flux follows 
from the equation of motion of a  scalar field (the dilaton) 
also present in that set-up \cite{quevedo}. 
In fact, for compact extra dimensions, the 4D Minkowski space is a unique solution with maximal 
symmetry \cite{gibbons}.  
The study in 6D gauged supergravity was generalized 
to axially symmetric solutions with a non-trivial 
warp factor \cite{gibbons,tasinato0,tasinato}.

%%%% Flux quantization problem  %%%%
Unfortunately, in both non-supersymmetric and supersymmetric setups, 
fine-tuning of the cosmological constant is not completely removed. 
It was noticed in \cite{navarro,quevedo,gibbons,fine-tuning} 
that taking into account the Dirac quantization condition for the flux 
results in another constraint, which relates different brane tensions 
to the discrete (monopole) number characterizing the flux configuration. 
As a consequence, 
stable background solutions can exist only for specifically chosen 
values of the brane tensions. 
In relation to the question of self-tuning,
the cosmology on a codimension-two brane in 6D flux models
has been studied in Refs.~\cite{cline} which come to the same conclusion.
It should be noted that this problem can be circumvented in 6D sigma models 
coupled to gravity. 
In such framework, the flux quantization constraint is absent and interesting 
brane solutions with  self-tuning features can be found \cite{sigma}.

%%%% New warped solutions %%%%
In this paper we extend the previous studies and present general warped solutions of 6D Salam-Sezgin supergravity. 
The dependence of the solution on the higher dimensional coordinates is given in terms of an arbitrary holomorphic function $V(z)$.
The number of conical singularities, 
which are free of any curvature divergence, 
is determined by the zeroes and poles of order $|\alpha| \leq 1$ 
of $V(z)$. Therefore, if $V$ is single-valued, 
it has only simple zeroes or poles. 
For the holomorphic function with a simple zero of $\alpha=1$,  
we recover the known warped solutions on $S_2$ 
with axial symmetry in extra dimensions \cite{gibbons,tasinato0,tasinato}. 
On the other hand, choosing $V(z)$ with many simple zeroes and/or poles, 
we can incorporate  
warped solutions with more than two branes (also without axial symmetry). 
In this case, however, there also appears a brane with fixed tension. 
We give an example for the warped solution with multi-branes and derive
the 4D Planck mass and the generalized flux quantization in this case.  

The paper is organized as follows.
After presenting our setup in Section 2,
we show the general warped solution and analyze its properties 
in Section 3.
In Section 4, we discuss solutions with two branes in some detail.
Then, in Section 5, we move to the warped solution with multi-branes.
Section 6 contains our conclusions.
Finally, a detailed derivation of the general warped solution
is presented in the Appendix.

%%%%%%%%%%%%%%%%%%%%%%%%%%%%%%%%%%%%%%%%%%%%%%%%%%%%%%%%%%%%%%%%%%%%
\section{The model}
%%%%%%%%%%%%%%%%%%%%%%%%%%%%%%%%%%%%%%%%%%%%%%%%%%%%%%%%%%%%%%%%%%%%

We consider the relevant\footnote{We have dropped the 2-form antisymmetric tensor field from the action, as it does not play any role in our solution.} 
bosonic action of the 6D gauged supergravity \cite{6dsugra1,6dsugra2} 
\begin{align}
  S_{\rm bulk}=\int d^6 X\sqrt{-G}\left[\frac{1}{2}R
  -\frac{1}{2}(\partial_M\Phi)^2-\frac{1}{4}e^{-\Phi}F_{MN}F^{MN}
  -2g^2\,e^{\Phi}\right],
\end{align}
supplemented with the 3-brane action:
\begin{align}
  S_{\rm brane}&=-\sum_i\int d^4x_i\sqrt{-g_i}\,\Lambda_i \,  \\
 &=\int d^6 X\sqrt{-G}\,\sum_i{\cal L}_{4,i}
\end{align}
where a distributional brane energy density is 
\begin{equation}
{\cal L}_{4,i}=-\int d^4x_i\sqrt{\frac{-g_i}{-G}}\,\Lambda_i\delta^6(X-X(x_i)).
\end{equation}
Here $\Lambda_i$ is the tension and $g_{i,\mu\nu}$ is the metric pulled back
to the brane worldvolume. 
The real scalar field $\Phi$ is called the dilaton.
The $U(1)$ field strength is defined as $F_{MN}=\partial_M A_N-\partial_N A_M$ and the gauge coupling is denoted by $g$.
The 6D fundamental scale has been suppressed, $M_6 = 1$.

The variation of the above action leads to the field equations 
\begin{align}
  \partial_M\left(\sqrt{-G}\,e^{-\Phi}F^{MN}\right)&=0, \label{eqf}\\
  \frac{1}{\sqrt{-G}}\,\partial_M\left(\sqrt{-G}\,\partial^M\Phi\right)&=
  -\frac{1}{4}e^{-\Phi}F_{MN}F^{MN}+2g^2e^{\Phi},\label{eqphi} 
\end{align}
 and 
the Einstein equations 
\begin{align}
  \begin{split}
    R_{MN}&=\partial_M\Phi\partial_N\Phi+g^2e^{\Phi}G_{MN}
    +e^{-\Phi}\left(F_{MP}F_N\,^P-\frac{1}{8}G_{MN}F_{PQ}F^{PQ}\right) \\ 
    &\quad +{\hat T}^b_{MN}
    \label{eqein} 
  \end{split} 
\end{align}
where ${\hat T}^b_{MN}$ is the brane contribution.

We look for a background solution that is  maximally symmetric 
in four dimensions. We  take the general warped ansatz
\begin{align}
  \label{e.wa}
  ds^2& =  W^2(y){\tilde g}_{\mu\nu}(x)dx^\mu dx^\nu+{\hat g}_{mn}(y)dy^m dy^n, \\
  F_{mn}&=\sqrt{\hat g}\,\epsilon_{mn}F(y),\\
  \Phi&=\Phi(y),
\end{align}
and other field components are assumed to have zero vevs. 
Greek letters ($\mu,\nu=0,\ldots,3$) denote the 4D coordinates, Roman letters ($m,n=5,6$) denote the
extra dimensional coordinates, ${\tilde g}_{\mu\nu}(x)$ is a metric of a 4D maximally symmetric
spacetime for which $R_{\mu\nu}({\tilde g})=3\lambda {\tilde g}_{\mu\nu}$, and $\epsilon_{mn}$ is
the Levi-Civita symbol. For this metric ansatz, nonzero Ricci tensor components are given by 
\begin{align}
  R_{\mu\nu}(G)&=\left(3\lambda-\frac{1}{4}W^{-2}D_m D^m W^4\right)
  {\tilde g}_{\mu\nu}\,,\\
  R_{mn}(G)&=R_{mn}({\hat g})-4W^{-1}D_m D_n W\,,
\end{align}
and the brane contribution to the Einstein equation is given by
\begin{equation}
{\hat T}^b_{MN}=-\sum_i\frac{\Lambda_i}{\sqrt{\hat g}}\, 
(g_{i,\mu\nu}\,\delta^\mu_M\delta^\nu_N-G_{MN})\delta^2(y-y_i)
\end{equation}
where $y_i$ are brane positions and $g_{i,\mu\nu}=W^2(y_i){\tilde g}_{\mu\nu}$.
Therefore, the field equations  
and the Einstein equations  become
\begin{align}
  \epsilon^{mn}\partial_m(W^4e^{-\Phi}F(y))&=0\,, \label{eq0} \\
  W^{-4}D_m(W^4D^m\Phi)&=
  -\bigg(\frac{1}{2}F^2e^{-\Phi}-2g^2e^{\Phi}\bigg)\,,\label{eq1}\\
  3\lambda-\frac{1}{4}W^{-2}D_m D^m W^4 
  &=-\bigg(\frac{1}{4}F^2e^{-\Phi}-g^2e^{\Phi}\bigg)W^2\,,
  \label{eq2}\\
  \begin{split}
    R_{mn}({\hat g})- 4 W^{-1}D_m D_n W&=\partial_m\Phi\partial_n\Phi
    +\bigg(\frac{3}{4}F^2e^{-\Phi}+g^2e^{\Phi}\bigg){\hat g}_{mn} \\
    &\quad 
    +\sum_i\frac{1}{\sqrt{\hat g}}\Lambda_i\,{\hat g}_{mn}\delta^2(y-y_i)\,.
    \label{eq3a}
  \end{split}
\end{align}

%%%%%%%%%%%%%%%%%%%%%%%%%%%%%%%%%%%%%%%%%%%%%%%%%%%%%%%%%%%%%%%%%%%%
\section{The general warped solution}
%%%%%%%%%%%%%%%%%%%%%%%%%%%%%%%%%%%%%%%%%%%%%%%%%%%%%%%%%%%%%%%%%%%%%%%%

In this section we present the general background solution of the 6D Salam-Sezgin supergravity with maximally symmetric four-dimensions and conical branes. 
Here we give only the final results, and postpone technical details 
of the derivation  until the Appendix.

We parameterize the extra dimensions 
with a complex coordinate $z = y_5 + i y_6$.
As shown in the Appendix, the compactness of extra dimensions requires 
the solution of the form (\ref{e.wa})  
only with the flat 4D metric, ${\tilde g}_{\mu\nu} = \eta_{\mu\nu}$.
The solution is fixed up to an arbitrary holomorphic function $V(z)$ and four real integration
constants $v$, $\Phi_0$, $f$ and $\zeta_0$. 
The warp factor is given in an implicit form as 
\begin{align}
  \label{e.swf}
  \frac{\left(W^4(\zeta) - W_-^4\right)^{W_-^4}}{\left(W_+^4 - W^4(\zeta)\right)^{W_+^4}} 
=  \exp\left\{2 \gamma^2 (W_+^4 - W_-^4)(\zeta-\zeta_0)\right\}
\end{align}
where 
\begin{equation}
  \zeta(z) = \frac{1}{2} (\xi+{\bar\xi})\,, \quad
  \xi=\int^z \frac{d\omega}{V(\omega)}\,, \label{zeta}
\end{equation}
and  
\begin{equation}
W_\pm^4 = v \pm \sqrt{v^2 - \frac{f^2}{4 g^2} }, \quad \gamma^2 = \frac{1}{4} g^2  e^{\Phi_0}\,.  \label{wbound} 
\end{equation}
The metric of the extra dimension is expressed 
in a conformally flat form as  
\begin{align}
  \label{e.sv}
  ds_2^2 &= e^{2K(z,\bar{z})} dz d \bar{z}\, , \qquad K =\frac{1}{2}\ln\left[\frac{1}{2|V(z)|^2} \frac{P(W)}{W^2}\right]
\end{align}
with
\begin{align}
  P(W)=\frac{1}{2}\gamma^2W^{-4}(W^4_+-W^4)(W^4-W^4_-).
\end{align}
The dilaton and the gauge flux are related to the warp factor via
\begin{align}
  \label{e.sd}
  \Phi&=\Phi_0- 2\ln W\,, & F(y)= f e^{\Phi_0}W^{-6}.
\end{align}
  
Let us now discuss important properties of the solution.
Note first that the warp factor is constrained \footnote{We note that 
in Romans supergravity \cite{Romans}, the scalar potential is negative
definite with $g^2\rightarrow -g^2$ in our setup.
In this case, there exists only one real root of $P(W)$ 
so the value of $W$ is not bounded.
Thus, for a finite 4D Planck mass, 
the space of extra dimensions must be terminated 
at a 4-brane \cite{tasinato0}.} to lie in the range  $W_- \leq W \leq W_+$.
This is the consequence of eq.~(\ref{e.sv}) and the fact that the extra dimensions 
are space-like. 
Furthermore, reality of the warp factor yields the constraint on the integration constants, $v >0$, $v^2 > f^2/(4 g^2)$.  
The extrema of the warp factor $W = W_{\pm}$ correspond to singular points of the metric. 
From eq.~(\ref{e.swf}) we find that $W = W_\pm$ for $\zeta =\pm \infty$.   

The allowed form of the holomorphic function is also constrained.
Suppose that $V(z)$ at $z =z_i$ is approximated by
$V(z) \approx  c_i^{-1} (z-z_i)^{\alpha_i}$ with $\alpha_i$ a real number. 
For $\alpha_i>1$, $\zeta$ and the warp factor 
are discontinuous around $z_i$. Moreover, for $\alpha_i<-1$, 
there exists a curvature singularity at $z_i$ as shown in the Appendix. 
Thus, the order of $V$ in local expansion is restricted to 
$-1\leq \alpha_i\leq 1$. Therefore,  we can conclude 
that a single-valued $V$ can have only simple zeroes or poles. 

Let us first consider the single-valued $V$ of the form
\begin{align}
  \label{e.zeroes}
  V(z) \approx c_i^{-1} (z-z_i) \qquad \Rightarrow \qquad \zeta \approx
\frac{1}{2}c_i \log |z-z_i|^2 \, ,
\end{align}
where $c_i$ must be real for the warp factor to be single-valued around $z_i$.
In this case, it follows that $\zeta \to \pm \infty$ for $z \to z_i$, 
so zeroes of $V(z)$ correspond to extrema of the warp factor 
and singular points of the 2D metric. 
That is, for $z \to z_i$, $W \to W_+$ if $c_i < 0$, 
and $W \to W_-$ for $c_i > 0$. 

For the simple zeroes of $V(z)$ of the form (\ref{e.zeroes}) 
we encounter conical singularities.  
Indeed, from eqs. (\ref{e.swf}) and (\ref{e.sv}), 
we find that the 2D metric behaves as 
\begin{align}
  K\approx  (\beta_\pm-1)\ln|z - z_i|, \qquad \qquad 
  \beta_\pm = \gamma^2|c_i|W^{-4}_\pm\left(W^4_+ -W^4_-\right).
\end{align}
Changing the coordinates to $\rho = |z - z_i|^{\beta_\pm}/\beta_\pm$, 
$\phi = {\rm Arg} (z - z_i)$, the 2D metric near $z = z_i$ becomes
$ds_2^2 \approx d\rho^2 + \beta_\pm^2 \rho^2 d\phi^2$. 
This is a flat 2D metric with a deficit angle $2 \pi (1-\beta_\pm)$.
Near $z = z_i$ only eq. (\ref{eq3a}) contains a singular term and it reduces to
$\partial {\bar\partial}K =  - \frac{\Lambda_i}{2} \delta^2(z-z_i)$.
Using the formula $\partial {\bar\partial} \log |z|^2 = 2 \pi \delta^2(z)$ we find that the deficit
angle should be  matched to the brane tensions as  
\begin{align}
  \Lambda_i=2\pi\left[1-\gamma^2|c_i|W^{-4}_\pm\left(W^4_+ -W^4_-\right)\right].
\end{align}
We see that the brane tensions are constrained 
to $\Lambda_i < 2 \pi$ (i.e., the deficit angle cannot exceed $2 \pi$).
If $V$ has more than two simple zeroes,
there also appears a brane with fixed tension. 

When we have a simple pole of the form $V(z)\sim c_i^{-1}(z-z_i)^{-1}$,
the deficit angle around $z_i$ is fixed to $-2\pi$ 
so the corresponding brane tension only takes a fixed value. 
On the other hand, 
when $V\approx c_i^{-1}(z-z_i)^{\alpha_i}$ with $|\alpha_i|<1$,
a brane with nontrivial tension could be located at $z_i$. 
However, since $V$ is not single-valued \footnote{We note that the warped solution does not have
a symmetry for accommodating a $SU(2)$ monodromy in $V$,
unlike the un-warped case \cite{redi}.} around $z_i$, the warp factor 
would not be well-defined around $z_i$.

The metric solution can be brought to an  axially symmetric form, 
which is a consequence of the existence of a Killing vector 
for the warped solution \cite{gibbons}.  
In order to see this we can perform the following locally well-defined change of coordinates:
\begin{align}
  d\zeta=\frac{1}{2}\left(\frac{dz}{V(z)} +\frac{d{\bar z}}{V({\bar z})}\right), \label{mapreal} \qquad 
  d\theta=\frac{1}{2i}\left(\frac{dz}{V(z)} -\frac{d{\bar z}}{V({\bar z})}\right).
\end{align}
In these new coordinates the metric solution is given by
\begin{subequations}
\begin{align}
  ds^2&=W^2 ds^2_4 +  \frac{P(W)}{2W^2} (d\zeta^2+d\theta^2) \label{sym1}\\
  &=W^2 ds^2_4 +\frac{W^4}{2P(W)}dW^2+\frac{P(W)}{2W^2}d\theta^2 \label{sym2}
\end{align}
\end{subequations}
where use is made of eq.~(\ref{master1}) in the second line.
Since the warp factor $W$ is a function of $\zeta$ only, the metric becomes 
independent of the angular variable $\theta$ in this local coordinate patch. 
If the mapping $(z,{\bar z}) \to (\zeta,\theta)$ is single-valued, then the change of coordinates (\ref{mapreal}) is globally well-defined.  
In this case, we obtain a warped solution with global axial symmetry,  
which will be discussed in the next section.
On the other hand, if the $(\zeta,\theta)$ coordinates do not cover  
the whole $z$ plane, the metric does not need to be axially symmetric.

%%%%%%%%%%%%%%%%%%%%%%%%%%%%%%%%%%%%%%%%%%%%%%%%%%%%%%%%%%%%%%%%%%%%
\section{Regular warped solutions with axial symmetry}
%%%%%%%%%%%%%%%%%%%%%%%%%%%%%%%%%%%%%%%%%%%%%%%%%%%%%%%%%%%%%%%%%%%%

In this section, as a particular solution in our formalism, 
we will recover the known warped solutions with axial
symmetry of extra dimensions in \cite{gibbons,tasinato0,tasinato} 
and discuss the flux quantization 
condition. Moreover, we will show a consistent procedure of taking
the limit that the warp factor becomes constant.

\subsection{The solution}

Now let us take a simple ansatz 
for $V(z)$ as 
\begin{align}
  V(z)=\frac{z}{c}\,, 
\end{align} 
with $c$ a complex number.
In this case, following the argument in the previous
section, there will appear a conical brane with nonzero tension at $z=0$. 
Then, the change of coordinates (\ref{mapreal}) become 
\begin{align}
  \zeta&=\frac{1}{2}\left(c\ln z+\bar{c}\ln {\bar z}\right), \label{t1} &\theta&=\frac{1}{2i}\left(c\ln
  z-\bar{c}\ln {\bar z}\right).
\end{align}
For the map $\zeta(z,{\bar z})$ to be single-valued around $z=0$,
we note that $c$ must be real \footnote{Also let us take a negative 
$c$. Taking a positive $c$ will not change physics.}. 
Thus, for this choice of $V(z)$,
there exists an axial symmetry of the metric and 
the change of coordinates is globally well-defined.
Since $\zeta$ is a log function of $|z|$, 
there will also appear another conical brane at $z=\infty$.

In this case, by redefining the variable in the metric form (\ref{sym1})
or (\ref{sym2}) as  
\begin{align}
  d\eta&=\frac{1}{|c|}\frac{dW}{WP(W)}=\frac{1}{|c|}W^{-4}d\zeta \label{eta}
\end{align} 
and inserting $\psi\equiv (\ln z-\ln{\bar z})/(2i)=-\theta/|c|$ in eq.~(\ref{t1}),
we can find the explicit form of the metric solution 
in the new coordinate as
\begin{align}
  ds^2=W^2\eta_{\mu\nu}dx^\mu dx^\nu
  +a^2W^8 d\eta^2+a^2 d\psi^2 \label{wsol}
\end{align}
where
\begin{align}
  W^4&=\frac{1}{2}\left(W^4_+ +W^4_-\right)
  +\frac{1}{2}\left(W^4_+-W^4_-\right)\tanh\left[\left(W^4_+-W^4_-\right)\gamma^2|c|\eta\right], \label{warp}\\ 
  \begin{split}
  a^2&=\frac{1}{2}|c|^2\frac{P(W)}{W^2}\\
	&=\frac{1}{16}\gamma^2|c|^2\left(W^4_+-W^4_-\right)^2 W^{-6}
  \cosh^{-2}\left[\left(W^4_+-W^4_-\right)\gamma^2|c|\eta\right].
  \end{split}
\end{align}

In order to see how the brane tensions are matched in the Einstein equation (\ref{zbz}), 
let us consider the asymptotic limits of the metric 
at two conical singularities.
For $\eta\rightarrow\pm\infty$, we get the warp factor as 
\begin{align}
  W^4\rightarrow W^4_\pm\mp\left(W^4_+-W^4_-\right)\exp\left\{\mp
    2\left(W^4_+-W^4_-\right)\gamma^2|c|\eta\right\}\,, 
\end{align}
as well as 
\begin{align}
  a^2\rightarrow \frac{1}{16}\gamma^2|c|^2 W^{-6}_\pm \left(W^4_+-W^4_-\right)^2 \exp\left\{\mp
    2\left(W^4_+-W^4_-\right)\gamma^2|c|\eta\right\}\,. 
\end{align}
Then, let us make a change of coordinate around each singularity 
by $d\rho_\pm=\mp aW^4 d\eta$, which goes to 
\begin{align}
  d\rho_\pm=\mp \frac{1}{2}\gamma |c|W_\pm \left(W^4_+-W^4_-\right) 
  \exp\left\{\mp \left(W^4_+-W^4_-\right)\gamma^2|c|\eta\right\} \,d\eta.
\end{align}
Consequently, the 2d metric goes to
\begin{align}
  ds^2_2&\rightarrow d\rho^2_\pm+\beta^2_\pm \rho^2_\pm d\psi^2\,,\qquad \beta_\pm\equiv \gamma^2|c|
  W^{-4}_\pm \left(W^4_+-W^4_-\right)\,. 
\end{align}

Therefore, we find the brane tensions located at $\eta=\eta_\pm$ to be
\begin{align}
  \Lambda_\pm=2\pi (1-\beta_\pm)
  =2\pi \left[1-\gamma^2|c|W^{-4}_\pm \left(W^4_+-W^4_-\right)\right].\label{brane}
\end{align}

Now let us look at the brane conditions more carefully to find the relations
between brane tensions explicitly.
By eliminating $\frac{W^4_+}{W^4_-}$ in the two brane conditions,
we can determine $e^{\Phi_0}|c|$  in terms of the brane tensions as
\begin{align}
  e^{\Phi_0}|c|=4 g^{-2}\,
  \frac{2\pi}{\Lambda_+-\Lambda_-}
  \bigg(1-\frac{\Lambda_+}{2\pi}\bigg)\bigg(1-\frac{\Lambda_-}{2\pi}\bigg).
  \label{tune0}
\end{align}
On the other hand, the remaining brane condition gives $\frac{W^4_+}{W^4_-}$ as
\begin{align}
  \frac{W^4_+}{W^4_-}=\frac{2\pi-\Lambda_-}{2\pi-\Lambda_+}
  \equiv \kappa>1\,, \label{tune1}
\end{align}
which gives rise to a relation between $f$ and $v$, 
\begin{align}
  \frac{f}{v}=\pm 4g\frac{\sqrt{\kappa}}{(\kappa+1)^2}\,. \label{tune2}
\end{align} 

Let us now consider the Planck mass for the warped solution with two branes.
Making a dimensional reduction of the 6D Einstein term 
for the metric form (\ref{sym2}),
we can read the 4D Planck mass as
\begin{align}
    M^2_\text{P}&=\frac{1}{2}M^4_6 \int^{W_+}_{W_-} W^3 dW \cdot\int d\theta\,.  \label{planck}
\end{align}
Then, by using eq.~(\ref{t1}), the Planck mass is shown to be finite as
\begin{align}
  \begin{split}
    M^2_\text{P}&=\frac{1}{4}\pi M^4_6 \left(W^4_+-W^4_-\right)|c| \\ 
    &=\frac{1}{2}\pi M^4_6 v|c| \sqrt{1-\left(\frac{f}{2gv}\right)^2} \label{plancka}
  \end{split}
\end{align}
where use is made of eq.~(\ref{wbound}) in the second line.

\subsection{Flux quantization}

Now let us consider the constraint coming from the flux quantization.
For compact dimensions, the gauge flux is quantized in the presence
of charged particles \cite{navarro,quevedo,gibbons,fine-tuning} as
\begin{align}
 \int\limits_{{\cal M}_2} F_2=\frac{2\pi n}{g}\label{fluxq}
\end{align}
with ${\cal M}_2$ being a 2D compact manifold and $n$ being an integer number.
For the metric form (\ref{sym2}), the flux is given by
\begin{align}
  \begin{split}
    F_{W\theta}&=\sqrt{\hat g}\,\epsilon_{W\theta}fe^{\Phi_0}W^{-6}=\frac{1}{2}\epsilon_{W\theta}\, fe^{\Phi_0}W^{-5}.
  \end{split}
\end{align}
Thus, the quantization condition becomes
\begin{align}
  \frac{1}{8}f e^{\Phi_0}\left(\frac{W^4_+-W^4_-}{W^4_+W^4_-}\right)
  \cdot\int d\theta=\frac{2\pi n}{g}.\label{quantum}
\end{align}
From eqs.~(\ref{quantum}) and (\ref{t1}),
the flux quantization can be rewritten as
\begin{align}
  \frac{W^4_+-W^4_-}{W^4_+W^4_-}\, f
  =\frac{8n}{g}\left(e^{\Phi_0}|c|\right)^{-1}.\label{quantuma}
\end{align}
Then, by using eqs.~(\ref{brane}) and (\ref{wbound}), 
we can rewrite the above equation as
\begin{align}
  \frac{\Lambda_+-\Lambda_-}{2\pi}=\frac{4n^2}{g^2}
  \left(e^{\Phi_0}|c|\right)^{-1}. \label{tune0a}
\end{align}
Therefore, with eq.~(\ref{tune0}), 
the flux quantization condition only gives a fine-tuning 
condition between the brane tensions as
\begin{align}
  \bigg(1-\frac{\Lambda_+}{2\pi}\bigg)\bigg(1-\frac{\Lambda_-}{2\pi}\bigg)=n^2.
\end{align}

Even after the flux quantization, we have fixed only 
one combination of parameters, $e^{\Phi_0}|c|$ via eq.~(\ref{tune0})
or (\ref{tune0a}),
and there is one relation between the parameters $f$ and $v$
from eq.~(\ref{tune2}). However, since $c$ can be absorbed by a rescaling of extra
coordinates, $\eta$ and $\psi$ in eq.~(\ref{wsol}), 
there appears only one undetermined modulus. However, as will be shown in
the next subsection, we find that it is necessary 
to keep $c$ in the solution explicit
to take the limit of the warp factor being constant 
while maintaining finite values of parameters of the resulting 
un-warped solution.

\subsection{The limit to the un-warped solution}

Let us consider a particular limit of the warp
solution that the warp factor becomes constant. 
For this purpose, let us find the consistent limiting procedure of parameters
in the solution. From the warped solution (\ref{wsol}) with eq.~(\ref{wbound}), 
we can see that the warp factor becomes constant for
\begin{align}
(W^4_+-W^4_-)\rightarrow 0^+ \ \ {\rm or} \ \ \frac{f}{v}\rightarrow \pm 2g.
\end{align}
In this limit, we take $|c|$ to infinity while keeping  
the following quantity finite 
\begin{align}
  k\equiv |c|\left(W^4_+-W^4_-\right).
\end{align}
This limit is also consistent with the relation of 
the holomorphic function $V$ to the warp factor as in eq.~(\ref{vdef}).
This is also necessary for maintaining the finite Planck mass 
in eq.~(\ref{plancka}) even for $\left(W^4_+-W^4_-\right)\rightarrow 0^+$.
Then, in this limit, the warped solution (\ref{wsol}) becomes
\begin{align}
  ds^2\rightarrow W^2_+\eta_{\mu\nu}dx^\mu dx^\nu +\frac{1}{16}\gamma^2W^2_+k^2
  \left[\frac{d\eta^2+W^{-8}_+d\psi^2}{\cosh^2(k\gamma^2\eta)}\right].
\end{align}
On making a change of coordinate as 
$d\rho=k\gamma^2 d\eta/\cosh(k\gamma^2\eta)$, the metric can be cast into the 
following form,
\begin{align}
  \begin{split}
    ds^2 &\rightarrow W^2_+\eta_{\mu\nu}dx^\mu dx^\nu
    +\frac{1}{16}\gamma^{-2}W^2_+ \left(d\rho^2+\beta^2\sin^2\rho
    d\psi^2\right),\\
    \beta &\equiv  k\gamma^2W^{-4}_+=\gamma^2|c| W^{-4}_+(W^4_+-W^4_-)\,. 
  \end{split}
\end{align}
Therefore, we get the smooth limit of our warped solution to the un-warped
solution where the finite quantity $k$ appears 
as a deficit angle. 
In fact, even though the new coordinate (\ref{zeta}) 
does not look well defined for a constant warp factor or $V=0$, 
from eqs.~(\ref{t1}) and (\ref{eta}), 
the change of the original complex coordinate 
to the final coordinate in eq.~(\ref{wsol}) is well defined for any value 
of $c$ as
\begin{align}
  d\eta=\frac{1}{2}W^{-4}\bigg(\frac{dz}{z}+\frac{d{\bar z}}{\bar z}\bigg).
\end{align} 
The resulting un-warped solution  
describes a sphere with two conical branes with equal tensions, 
\begin{align}
  \Lambda_+=\Lambda_-=2\pi(1-\beta).
\end{align}
This result is consistent with the limit of the brane conditions (\ref{brane}).

%%%%%%%%%%%%%%%%%%%%%%%%%%%%%%%%%%%%%%%%%%%%%%%%%%%%%%%%%%%%%%%%%%%

\section{Warped solutions with multi-branes}

%%%%%%%%%%%%%%%%%%%%%%%%%%%%%%%%%%%%%%%%%%%%%%%%%%%%%%%%%%%%%%%%%%%

In this section, we generalize the previous warped solution 
with two branes to the case with more than two branes. 
Moreover, we find the generalized flux quantization condition
and discover new un-warped solutions with multi-branes by taking 
the limit of the warp factor being constant.

\subsection{The solution}

Let us take a more general form of holomorphic function $V$, 
\begin{align}
V(z)=\frac{1}{c}\prod_{i=1}^N(z-z_i) \label{vsolb}
\end{align}
where $c$ and $z_i\,(i=1,\cdots, N)$ are complex numbers.
Then, from eq.~(\ref{zeta}), the new coordinate $\zeta$ becomes
\begin{align}
\zeta=\frac{1}{2}\sum_{i=1}^N
\left(c\int dz\prod_{i=1}^N\frac{1}{(z-z_i)}
+{\rm c.c.}\right)\,.
\end{align}
By using
\begin{align}
  \prod_{i=1}^N\frac{1}{(z-z_i)} &= \sum_{i=1}^N \left(\prod_{j\neq i}\frac{1}{(z_i-z_j)}\right) \frac{1}{z-z_i},
\end{align}
we can rewrite $\zeta$ as
%\begin{widetext}
\begin{align}
  \zeta=\frac{1}{2}\sum_{i=1}^N\left[
    c\prod_{j\neq i}\frac{1}{(z_i-z_j)}\ln(z-z_i)
    +\bar{c}\prod_{j\neq i}\frac{1}{({\bar z}_i-{\bar z}_j)}
    \ln({\bar z}-{\bar z}_i)\right].
\end{align}
%\end{widetext}
Then, for the single-valuedness of the warp factor $W(\zeta)$ at $z=z_i\,(i=1,\cdots,N)$, we must
require 
\begin{align}
  c\prod_{j\neq i}\frac{1}{(z_i-z_j)} & =\bar{c}\prod_{j\neq i}\frac{1}{({\bar z}_i-{\bar z}_j)}
  \quad \text{for all } i\, .
\end{align}
This means that all the $z_i$'s have to be aligned, $z_i = e^{i\phi}\left|z_i\right|$ with a common
phase $\phi$, while $c=e^{- i (N-1)\phi} \left|c\right|$. By a change of coordinates and a
redefinition of $c$, we can take all the $z_i$'s and $c$ to be real. 
In this case, $\zeta$ becomes
\begin{align}
  \zeta=\frac{1}{2}\sum_{i=1}^Na_i\ln|z-z_i|^2\label{tsol1} \,, \quad  \,a_i\equiv
  c\prod_{j\neq i}\frac{1}{(z_i-z_j)}\,. 
\end{align}

Now we can see that $z=z_i$ is mapped
to $\zeta=+\infty\;(-\infty)$ for $a_i<0\;(a_i>0)$,
finally mapped to $W=W_+(W_-)$.
On the other hand, because $\sum_{i=1}^N a_i=0$, 
we also find that $z=\infty$ is mapped to $\zeta=0$.

By following a similar analysis as in the previous section,
we can identify the brane tensions located at $z=z_i$ as
\begin{align}
  \Lambda^i_\pm=2\pi\left [1-\gamma^2|a_i|W^{-4}_\pm \left(W^4_+-W^4_-\right)\right],\quad
  a_i<0\,\; (a_i>0)\,.\label{brane22}
\end{align}
Moreover, 
we can also find the brane tension located at $z=\infty$ fixed as
\begin{align}
  \Lambda^\infty=2\pi(2-N).
  \label{inft21}
\end{align}
Therefore, 
the brane tension at $z=\infty$
is fixed by the number of zeroes of $V(z)$, $N$.
We find that eqs.~(\ref{brane22}) and (\ref{inft21}) for $N=2$
reproduce the two-brane solution.

Let us see the conditions between unfixed brane tensions explicitly. 
Suppose that $a_i<0$ for $i=1,\cdots,k$ and $a_i>0$ for $i=k+1,\cdots,N$.
Then, summing $\Lambda^i_+$ over $i=1,\cdots,k$ and
$\Lambda^i_-$ over $i=k+1,\cdots,N$ from eq.~(\ref{brane22}), 
we can get a similar condition as eq.~(\ref{tune1}),
\begin{align}
  \frac{W^4_+}{W^4_-}&= \frac{2\pi(N-k)-\sum_{i=k+1}^N\Lambda^i_-}{2\pi
    k-\sum_{i=1}^k\Lambda^i_+} \equiv {\tilde\kappa} > 1\,.\label{tune1a} 
\end{align}
This gives rise to the  relation between $f$ and $v$ as
\begin{align}
  \frac{f}{v}=\pm 4g \frac{\sqrt{\tilde\kappa}}{(\tilde\kappa+1)^2}\,. \label{tune2a} 
\end{align}
Moreover, eliminating $\frac{W^4_+}{W^4_-}$ in the brane
condition (\ref{brane22}), we can determine $e^{\Phi_0}|a_i|$
in terms of the brane tensions: for $i=1,\cdots,k$,
\begin{align}
  \begin{split}
    e^{\Phi_0}|a_i|&=
    4g^{-2}\left(N-2k+\frac{\sum_{j=1}^k\Lambda^j_+-\sum_{j=k+1}^N\Lambda^j_-}{2\pi}\right)^{-1}\\ 
    &\quad \mspace{90mu}\times \left(1-\frac{\Lambda^i_+}{2\pi}\right)
\left(N-k-\frac{\sum_{j=k+1}^N\Lambda^j_-}{2\pi}\right), 
  \label{tune00a} 
  \end{split}
\end{align}
and for $i=k+1,\cdots,N$,
\begin{align}
  \begin{split}
    e^{\Phi_0}|a_i|&=
    4g^{-2}\left(N-2k+\frac{\sum_{j=1}^k\Lambda^j_+-\sum_{j=k+1}^N\Lambda^j_-}{2\pi}\right)^{-1}\\
    &\quad\mspace{90mu}\times
    \left(1-\frac{\Lambda^i_-}{2\pi}\right)
    \left(k-\frac{\sum_{j=1}^k\Lambda^j_+}{2\pi}\right). \label{tune00b}
  \end{split}
\end{align}
Thus, from eqs.~(\ref{tune00a}) and (\ref{tune00b}),
we can see that $\sum_{i=1}^N a_i=0$ is satisfied.
Therefore, from eqs.~(\ref{tune2a}), (\ref{tune00a}) and (\ref{tune00b}),
the brane conditions only fix the following combinations
of parameters: $e^{\Phi_0}|a_i|\; (i=1,\cdots,N)$ and $f/v$. 
Although two of the brane positions can be always chosen at $z=(0,1)$ 
by the invariance of the warp factor, the overall constant of $V$ cannot be 
absorbed by a rescaling of extra coordinates while keeping the brane positions, 
unlike the case with two branes.  Nonetheless, there appears only one undetermined modulus as in the
warped solution with two branes.

Let us now consider the Planck mass for the warped solution with multi-branes.
With the metric solution (\ref{metricw}) with eq.~(\ref{vdef}), 
a dimensional reduction of the 6D Einstein term to 4D leads to the 
Planck mass as
\begin{align}
  M^2_\text{P}&= \frac{1}{8}M^4_6\int dz d{\bar z} \,
  \frac{{\bar\partial}W^4}{V}\,. \label{planckgen} 
\end{align}
Here we note that the divergence theorem in the complex plane is given by 
\begin{align}
  \int\limits_R dz d{\bar z}\,(\partial {\bar J}+{\bar\partial}J) &= i\oint\limits _{\partial
    R}({\bar J}d{\bar z}-Jdz) 
\end{align}
with $J$ a complex function and $R$ the integration surface with boundary $\partial R$.
Then, by regarding the surface of compact dimensions as the sum of multiple patches, and applying
the divergence theorem, eq.~(\ref{planckgen}) becomes
\begin{align}
  M^2_\text{P}=\frac{1}{4}\pi M^4_6\left(\sum_{i=1}^k|a_i|\right)
  \left(W^4_+-W^4_-\right). \label{planckmulti} 
\end{align}

\subsection{Generalized flux quantization}

Now let us consider the flux quantization condition for the warped  
solution with multi-branes.  For the metric (\ref{metricw}) with eq.~(\ref{vdef}) in the $z$ complex
coordinate, the flux is given by 
\begin{align}
  \begin{split}
    F_{z{\bar z}}&= \frac{1}{2}\epsilon_{z{\bar z}}f e^{\Phi_0}W^{-4}e^{2A} \\
    &= -\epsilon_{z{\bar z}}f e^{\Phi_0} \,\frac{{\bar\partial} (W^{-4})}{8V}\,.
  \end{split}
\end{align}
Then, from eq.~(\ref{fluxq}), the generalized flux quantization condition is 
\begin{align}
  -\frac{1}{8}f e^{\Phi_0}\int dz d{\bar z}\, \frac{{\bar\partial} (W^{-4})}{V}&=\frac{2\pi n}{g}. 
\end{align}
Making a similar computation of the complex integral as for the Planck mass
in the previous section, we obtain the flux quantization condition as
\begin{align}
  f e^{\Phi_0}\left(\sum_{i=1}^k|a_i|\right)(W^{-4}_+-W^{-4}_-&)=\frac{8n}{g}. 
\end{align}
Thus, taking the sum of eq.~(\ref{tune00a}),
the flux quantization condition corresponds
to a generalized version of
fine-tuning condition between brane tensions as
\begin{align}
  \left(k-\frac{\sum_{i=1}^k\Lambda^i_+}{2\pi}\right)
  \left(N-k-\frac{\sum_{i=k+1}^N\Lambda^i_-}{2\pi}\right) =n^2\,.
\end{align}

\subsection{The limit to the un-warped solution}

In this section, we generalize the procedure of taking the limit 
of the warp factor being constant,
from the case with two branes to the one with multi-branes.

From the explicit form of the warp factor (\ref{explicit}),
let us define a function $\chi$ as
\begin{align}
  e^{W^4_+ \chi}&\equiv (W^4-W^4_-)\,e^{2\gamma^2\zeta}\,, \\
  e^{W^4_- \chi}&\equiv (W^4_+-W^4)\,e^{2\gamma^2\zeta}
\end{align}
where we dropped the integration constant for simplicity.
Then, the warp factor can be written as
\begin{align}
  W^4=\frac{1}{2}(W^4_++W^4_-)+\frac{1}{2}\left(W^4_+-W^4_-\right)
  \tanh\left[\frac{1}{2}\left(W^4_+-W^4_-\right)\chi\right] 
\end{align}
with
\begin{align}
  e^{W^4_+\chi}+e^{W^4_-\chi}=\left(W^4_+-W^4_-\right)\,e^{2\gamma^2\zeta}.
\end{align}

Now let us take the limit that the warp factor becomes constant.
For \mbox{$\left(W^4_+-W^4_-\right)\rightarrow 0^+$}, 
the general warped solution becomes
\begin{align}
  ds^2\rightarrow W^2_+\eta_{\mu\nu}dx^\mu dx^\nu
  +\frac{1}{16}
\frac{\gamma^2W^{-6}_+\left(W^4_+-W^4_-\right)^2}{|V|^2\cosh^2\left[W^{-4}_+\left(W^4_+-W^4_-\right)\gamma^2\zeta\right]}\,dz
d{\bar z}\,. 
\end{align}
Then, we can rewrite the resulting un-warped solution as
\begin{equation}
    ds^2\rightarrow W^2_+\eta_{\mu\nu}dx^\mu dx^\nu
    +\frac{1}{4}\gamma^{-2}W^2_+\frac{|\partial\omega|^2}{(1+|\omega|^2)^2} \, dz d{\bar z}\,
\end{equation}
with
\begin{equation}
    \omega\equiv \exp
\left\{W^{-4}_+\left(W^4_+-W^4_-\right)\gamma^2\xi\right\}.
\end{equation}

For instance, given the ansatz of $V$ for multi-branes in eq.~(\ref{vsolb}), 
we find the corresponding holomorphic function $\omega$ in the un-warped 
solution as
\begin{align}
  \omega =\prod_{i=1}^N(z-z_i)^{\beta_i}\,,\qquad \beta_i\equiv \gamma^2 a_i W^{-4}_+\left(W^4_+-W^4_-\right)\,.
\end{align}
In this case, the appearing brane tensions in the un-warped solution are 
\begin{align}
  \begin{split}
    \Lambda^i &=2\pi \left(1-|\beta_i|\right),  \qquad i=1,\ldots,N, \\
    \Lambda^\infty& = \mspace{4mu}2\pi\left(2-N\right). \label{uwb}
  \end{split}
\end{align}
Thus, this result is consistent with the limit of the brane conditions, 
eqs.(\ref{brane22}) and (\ref{inft21}). 
Therefore, for $\left(W^4_+-W^4_-\right)\rightarrow 0^+$ and finite $\beta_i$'s,
we can obtain the smooth limit of the warped solution with multi-branes
to the un-warped solution. 

From eq.~(\ref{planckmulti}), in the limit of a constant warp factor, the Planck mass becomes 
\begin{align}
  M^2_\text{P}&\rightarrow \frac{1}{4}\pi M^4_6\left(\sum_{i=1}^k|\beta_i|\right)\gamma^{-2}
  \label{unplanck} 
\end{align}
which agrees with the result for the un-warped metric 
with brane tensions given in eq.~(\ref{uwb}) and the 2D curvature
of $16\gamma^2$.

%%%%%%%%%%%%%%%%%%%%%%%%%%%%%%%%%%%%%%%%%%%%%%%%%%%%%%%%%%%%%%%%%%%%

\section{Conclusion}

%%%%%%%%%%%%%%%%%%%%%%%%%%%%%%%%%%%%%%%%%%%%%%%%%%%%%%%%%%%%%%%%%%%%

We have presented the general regular warped solutions 
with 4D Minkowski spacetime 
in six-dimensional gauged supergravity. 
The explicit form of the solution is determined 
up to an arbitrary holomorphic function which is constrained only
for the single-valueness of the warp factor. 
In this formalism, we have seen that 
multiple conical branes can be embedded into the warped geometry 
(also without the axial symmetry of extra dimensions). 

For the holomorphic function with one simple zero,
we have recovered the known warped solution 
with two conical branes and discussed the flux quantization in this case. 
On the other hand, taking a more general form of the holomorphic function
with multiple simple zeroes, we found the warped solutions with more than two
branes and obtained the Planck mass and the flux quantization condition
in an explicit form by complex integrals. 
In this case, a brane with fixed tension is necessary. 

The presence of the undetermined holomorphic function 
enables us to accommodate  
the warped solution with extra dimensions of different geometry than $S_2$. 
In particular, if $V(z)$ is double periodic the solutions fit
in the torus geometry \cite{gaussbonnet}. 
We leave the detailed discussion on this possibility 
in a future publication \cite{fll}.

\vskip1cm

\begin{acknowledgments}
  It is a pleasure to thank A.~Falkowski for collaboration and discussion 
at various stages of this work. We would like to thank
W. Buchm\"uller for comments in improving the manuscript 
and G. Tasinato for discussion.   
We are also grateful to M.~Redi for correspondences.  
HML has benefited from the COSMO-05 workshop where this work was presented. 
\end{acknowledgments}

%%%%%%%%%%%%%%%%%%%%%%%%%%%%%%%%%%%%%%%%%%%%%%%%%%%%%%%%%%%%%%%%%%%%%%%%%%%%%%%%%%%%%%%%%%%%%%%%%%%
%%%%%%%%%%%%%%%%%%%%%%%%%%%%%%%%%%%%%%%%%%%%%%%%%%%%%%%%%%%%%%%%%%%%%%%%%%%%%%%%%%%%%%%%%%%%%%%%%%%

\appendix

\section*{Appendix: Derivation of the general warped solution}
\addcontentsline{toc}{section}{Appendix: Derivation of the general warped solution}
\def\theequation{\thesubsection-\arabic{equation}}
\def\thesubsection{A}
\setcounter{equation}{0}

%\subsection*{Derivation of the general warped solution}
\label{a.der}

We consider the equations of motion of the 6D Salam-Sezgin supergravity, 
eqs. (\ref{eq0}) to (\ref{eq3a}). 
Before finding an explicit solution let us study some general consequences 
of the compactness of extra dimensions and the regularity of the solution.
From eqs.~(\ref{eq1}) and (\ref{eq2}), we get 
\begin{align}
  D_m\left(W^4D^m(\Phi+2\ln W)\right)=6\lambda W^2.
  \label{condition}
\end{align}
The left-hand side vanishes when integrated over the compact extra dimensions, which implies $\lambda=0$. 
Thus, 4D flat space is a unique maximally symmetric solution \cite{gibbons}. 
Moreover, let us assume that the compact extra dimensions are free of singularities and $\Phi$ and $W$ are regular in extra dimensions.
Then, by multiplying eq.~(\ref{condition}) with $\lambda=0$ by $(\Phi+2\ln W)$
and integrating by parts, we obtain 
\begin{align}
  \Phi=\Phi_0-2\ln W \label{scalar}
\end{align}
with $\Phi_0$ being a real constant.  

Next, the equation of motion for  the gauge field (\ref{eq0}) 
immediately leads to the solution for the field strength: 
\begin{align}
  F(y)=f e^{\Phi}W^{-4}=f e^{\Phi_0}W^{-6}\label{fst}
\end{align}
with $f$ being a real constant.

Now let us turn to the Einstein equations. 
Using $R_{mn}({\hat g})=K(y){\hat g}_{mn}$ with $K$ being the Gauss curvature and taking the
relations (\ref{scalar}) and (\ref{fst}), the $R_{mn}$ equation (\ref{eq3a}) becomes  
\begin{align}
  \begin{split}
  K{\hat g}_{mn}-2W^{-2}D_m D_n W^2&= e^{\Phi_0}\bigg(\frac{3}{4}f^2W^{-10}+g^2W^{-2}\bigg){\hat
    g}_{mn}\\
  &\quad +\sum_i\frac{1}{\sqrt{\hat g}}\Lambda_i\,{\hat g}_{mn}
  \,\delta^2(y-y_i)\,.\label{gausscurv} 
  \end{split}
\end{align}
Therefore, the trace-free part gives
\begin{align}
  D_m D_n W^2=\frac{1}{2}D^2W^2\, {\hat g}_{mn}.
\end{align}
This implies the existence of a Killing vector \cite{gibbons},
\begin{align}
  V_m=\sqrt{\hat g}\,\epsilon_{mn}D^n W^2
\end{align}
satisfying $D_{(m}V_{n)}=0$. 

By integrating the trace of eq.~(\ref{gausscurv}) over the extra dimensions, we get the Euler number as
%\begin{widetext}
\begin{align}
  \chi=\frac{1}{2\pi}\int d^2 y\sqrt{\hat g}
  \left[W^{-2}D^2W^2+e^{\Phi_0}\Big(\frac{3}{4}f^2W^{-10}+g^2 W^{-2}\Big)\right]
  +\frac{1}{2\pi}\sum_i\Lambda_i.
\end{align}
%\end{widetext}

Now let us introduce a complex coordinate $z = y_5 + i y_6$ 
and take the following ansatz for the metric: 
\begin{align}
  ds^2=W^2(z,{\bar z})\left(\eta_{\mu\nu}dx^\mu dx^\nu +e^{2A(z,{\bar z})}dz 
    d{\bar z}\right)\label{metricw}
\end{align}
where $\eta_{\mu\nu}$ is the 4D flat metric.
Inserting the above ansatz into (\ref{eq2}) and (\ref{eq3a})  with eqs.~(\ref{scalar}) and (\ref{fst}), 
the $R_{\mu\nu}$, $R_{zz}$, and $R_{z{\bar z}}$ equations become, 
respectively,
\begin{align}
  -4e^{-2A}(4\partial B{\bar\partial}B+\partial{\bar\partial}B)
  &=e^{\Phi_0}\bigg(g^2-\frac{1}{4}f^2e^{-8B}\bigg), \label{mn}\\ 
  -4\partial^2 B+4(\partial B)^2+8\partial B \partial A&=4(\partial
B)^2, \label{zz}\\ 
  \begin{split}
  -6\partial{\bar\partial}B-4\partial B{\bar\partial}B-2\partial{\bar\partial}A &=4\partial
  B{\bar\partial}B
+\frac{1}{2}e^{2A}e^{\Phi_0}\bigg(g^2+\frac{3}{4}f^2e^{-8B}\bigg)\\
	&\quad  +\sum_i
  \Lambda_i\delta^2(z-z_i) \label{zbz} 
  \end{split}
\end{align} 
with $B\equiv \ln W$, $\partial\equiv\frac{\partial}{\partial z}$ and 
${\bar \partial}\equiv\frac{\partial}{\partial {\bar z}}$.

In order to find a solution to the bulk equations we first rewrite eq.~(\ref{zz}) as 
\begin{align}
  e^{2A}\,{\bar\partial}(e^{-2A}{\bar\partial}B)=0.\label{zzagain}
\end{align}
This can be easily solved up to  an arbitrary  holomorphic function $V(z)$, 
\begin{align}
  V(z) = e^{-2A}{\bar\partial}B. \label{vdef}
\end{align}
For constant $B$ (i.e.~$V=0$), from eq.~(\ref{mn}) we find the condition $f^2=4g^2$.
% and we only have to solve eq.~(\ref{zbz}).
This choice corresponds to un-warped solutions that were considered in  ref. \cite{quevedo}
and (with multiple branes) in  ref. \cite{redi}.
Here we will focus on the warped solution. 
For $V\neq 0$ we can multiply both sides of eq.~(\ref{mn}) by $V$ and  integrate over ${\bar z}$ to get 
\begin{align}
  V\partial e^{4B}=-\frac{1}{4}e^{\Phi_0}g^2\left(e^{4B}+\frac{f^2}{4g^2} e^{-4B}-2v(z)\right)
  \label{master0} 
\end{align}  
with $v(z)$ being a holomorphic function. 
Let us now introduce a new holomorphic variable\footnote{There was a similar 
  approach to finding a warped solution in 6D Einstein gravity 
  with a nonzero bulk cosmological constant \cite{chodos}. In their case, 
  only 4D dS space solution is allowed.}
\begin{align}
  \xi=\int^z \frac{d\omega}{V(\omega)}.\label{newxi}
\end{align}
Then, we can rewrite the equation (\ref{master0}) as
\begin{align}
  \frac{\partial}{\partial\xi}W^4=-\frac{1}{4}e^{\Phi_0}g^2
  \left(W^4+\frac{f^2}{4g^2}W^{-4}-2v(\xi)\right). \label{master}
\end{align}
Combining the reality of the warp factor $W$ and the holomorphicity of $v(\xi)$,  we find that $W$
is a function of real part of $\xi$ only and $v$ is a real constant.   
That is, for the real variable
\begin{align}
  d\zeta=\frac{1}{2}d\left(\xi+{\bar\xi}\right) =\frac{1}{2}\left(\frac{dz}{V(z)}+\text{c.c.}\right),\label{tdef} 
\end{align}
eq.~(\ref{master}) becomes
\begin{align}
  f(W)dW\equiv \frac{W^3}{P(W)}dW=d\zeta \label{master1}
\end{align}
where
\begin{align}
  P(W)&=\frac{1}{2}\gamma^2\left(-W^4-u^2W^{-4}+2v\right) \label{pdef}\,,\quad \gamma^2 \equiv
  \frac{1}{4}e^{\Phi_0}g^2\,,\quad u^2\equiv \frac{f^2}{4g^2}\, . 
\end{align}

Therefore, we find that the equation of motion is reduced to the simple ordinary differential
equation (\ref{master1}). This can be integrated easily to yield
\begin{align}
  \frac{\left(W^4(\zeta)-W^4_-\right)^{W^4_-}}
{\left(W^4_+-W^4(\zeta)\right)^{W^4_+}}
  &={\rm exp}\left\{2\left(W^4_+-W^4_-\right)\gamma^2(\zeta-\zeta_0)\right\} 
\label{explicit} 
\end{align}
with $\zeta_0$ an integration constant and 
\begin{align}
  W_\pm^4 = v \pm \sqrt{v^2 - u^2 }  \, .
\end{align} 
One can  also show that this solution satisfies 
the remaining equation (\ref{zbz}) away from the branes. 
This completes the derivation of the general solution, 
eqs.~(\ref{e.swf}) to (\ref{e.sd}).

Before closing  this section, let us comment on the possible curvature 
singularities of our general warped solutions.
The 6D Ricci scalar for our general warped solutions is given as
\begin{align}
  R=4\gamma^2W^{-6}(5W^4 - u^2 W^{-4} + 2 v).
\end{align}
Thus, the 6D Ricci scalar is finite, as $W$ is constrained  in the range $W_-\leq W\leq W_+$.
Furthermore, let us also consider a higher curvature invariant in the bulk such as
\begin{align}
  R_{MNPQ}R^{MNPQ}=-24W^{-4}(DW)^4-16W^{-2}(D_mD_n W)^2+4K^2 \label{hcurv}
\end{align}
with the Gaussian curvature from eq.~(\ref{gausscurv}) as
\begin{align}
  K=W^{-2}D^2 W^2+e^{\Phi_0}\left(\frac{3}{4}f^2W^{-10}+g^2W^{-2}\right).
\end{align}
By plugging into eq.~(\ref{hcurv}) the following quantities
\begin{align}
  (DW)^2 &= 2W^{-4}P(W), \\
  D^2W &= 2W^{-5}(WP'(W)-3P(W)), \\
  (D_mD_nW)^2 &= W^{-10}\left[8\left|-3\frac{V}{\overline V}P
    +2W^4{\bar\partial}V\right|^2+2(WP'(W)-3P(W))^2\right], 
\end{align}
we can see that ${\bar\partial}V$ must be finite 
for a regular higher curvature invariant. Certainly, $\bar{\partial} V=0$ for any regular point of
$V$. At singularities, however, $\bar{\partial} V$ can be different from zero.
Suppose $V$ has an expansion $V\sim (z-z_i)^{\alpha_i}$ around $z_i$ with $\alpha_i$ a real number.
We note that for $\alpha_i\neq -1$,
\begin{align}
  \begin{split}
    {\bar\partial} (z-z_i)^{\alpha_i}&=-\alpha_i(z-z_i)^{\alpha_i+1}
    \underbrace{\bar{\partial}\, 
      \frac{1}{z}}_{\mathclap{\bar{\partial}\partial 
        \ln\left|z\right|^2}}+\bar{\partial}(z-z_i)^{\alpha_i}\Big|_\text{BC} \\
    &= -2\pi \alpha_i (z-z_i)^{\alpha_i+1} 
    \delta^2\left(z-z_i\right)\\
    &\quad +\frac{i}{2}|z-z_i|^{\alpha_i-1}e^{-i\Arg(z-z_i)}(1-e^{2\pi i\alpha_i})\,
    \delta\left(\Arg(z-z_i)\right)
  \end{split}
\end{align}
where BC is the branch cut for $z_i$ with non-integer $\alpha_i$.
Thus we can distinguish several cases:     
\begin{itemize}
  \item
    For natural number $\alpha_i$, ${\bar\partial }V$ vanishes. 
   
  \item
   For non-integer number satisfying $\alpha_i>1$ or $-1<\alpha_i<1$, 
there is a 1D delta function singularity along the branch cut.

  \item
   For $\alpha_i=-1$, there is a 2D delta function singularity.

   \item
   For $\alpha_i<-1$, there is a curvature singularity.
  
\end{itemize}
For $\alpha_i>1$, $\zeta$ and hence the warp factor is
discontinuous around $z_i$.
The delta funtion singularities appearing in the second and third cases
are as singular as the conical singularities, in the sense that they 
can be regularized at the level of higher order curvature expansion. 
Therefore, in the end, we have to restrict ourselves to $|\alpha_i|\leq 1$.
\vfill

%%%%%%%%%%%%%%%%%%%%%%%%%%%%%%%%%%%%%%%%%%%%%%%%%%%%%%%%%%%%%%%%%%%%%%%
%
%   bibliography
%
%

\pagebreak

\end{document}